\documentclass[aps,twocolumn,showpacs,superscriptaddress,amsmath,amssymb]{revtex4}
\usepackage{graphicx}
\usepackage{dcolumn}
\usepackage{bm}

\begin{document}

\preprint{NUHEP-EXP/07-04}
\title{Electron - Dark Matter Scattering in an Evacuated Tube}

\author{Yonatan Kahn}
\email{y-kahn@northwestern.edu}
\author{Michael Schmitt}
\email{schmittm@lotus.phys.northwestern.edu}
\affiliation{Northwestern University, Department of Physics and Astronomy,
2145 Sheridan Road, Evanston, IL 60208, USA}
\newcommand{\keV}                {{\mathrm{keV}}}
\newcommand{\MeV}                {{\mathrm{MeV}}}
\newcommand{\GeV}                {{\mathrm{GeV}}}
\newcommand{\Mchi}               {M_{\chi}}
\newcommand{\Mchisq}             {M_{\chi}^2}
\newcommand{\MU}                 {M_{U}}
\newcommand{\MUsq}               {M_{U}^2}
\newcommand{\Cchi}               {C_\chi}
\newcommand{\Cchisq}             {C_\chi^2}
\newcommand{\fe}                 {f_e}
\newcommand{\fesq}               {f_e^2}
\newcommand{\fnu}                {f_\nu}
\newcommand{\fnusq}              {f_\nu^2}
\newcommand{\Beeann}             {B^{ee}_{\mathrm{ann}}}
\newcommand{\epem}               {e^+e^-}
\newcommand{\dmdensity}          {\rho = 0.3~{\mathrm{GeV}}/{\mathrm{cm}}^3}
\newcommand{\Ebeam}              {E_{\mathrm{beam}}}
\newcommand{\EbeamVAL}           {100~{\mathrm{MeV}}}
\newcommand{\Ibeam}              {I_{\mathrm{beam}}}
\newcommand{\IbeamVAL}           {50~{\mathrm{kA}}}
\newcommand{\Ltube}              {L_{\mathrm{tube}}}
\newcommand{\LtubeVAL}           {100~{\mathrm{m}}}
\newcommand{\Rtube}              {r_{\mathrm{tube}}}
\newcommand{\RtubeVAL}           {2.5~{\mathrm{cm}}}
\newcommand{\Router}             {R_{\mathrm{outer}}}
\newcommand{\RouterVAL}          {20~{\mathrm{cm}}}
\newcommand{\Sidist}             {d_{\mathrm{Si}}}
\newcommand{\SidistVAL}          {20~{\mathrm{cm}}}
\newcommand{\Sithick}            {100~\mu{\mathrm{m}}}
\newcommand{\calthick}           {30~{\mathrm{cm}}}
\newcommand{\BVAL}               {300~{\mathrm{G}}}
\newcommand{\uhv}                {p_{\mathrm{tube}}}
\newcommand{\uhvVAL}             {10^{-12}~{\mathrm{Torr}}}
\newcommand{\lessvac}            {p_{\mathrm{outer}}}
\newcommand{\lessvacVAL}         {10^{-6}~{\mathrm{Torr}}}
\newcommand{\Errsi}              {\sigma_{\mathrm{si}}}
\newcommand{\ErrsiVAL}           {50~\mu{\mathrm{m}}}
\newcommand{\ThetaminVAL}        {10^{\circ}}
\newcommand{\pb}                 {\mathrm{pb}}

\newcommand{\bi}   {\begin{itemize}}
\newcommand{\ei}   {\end{itemize}}
\newcommand{\be}   {\begin{enumerate}}
\newcommand{\ee}   {\end{enumerate}}
\newcommand{\bcen} {\begin{center}}
\newcommand{\ecen} {\end{center}}
\newcommand{\beq}  {\begin{equation}}
\newcommand{\eeq}  {\end{equation}}
\newcommand{\bdm}  {\begin{displaymath}}
\newcommand{\edm}  {\end{displaymath}}

\newcommand{\etal} {{\em et al.}}
\newcommand{\ie} {{\em i.e.}}
\newcommand{\eg} {{\em e.g.}}

\date{June 15, 2008}

\begin{abstract}
The light dark matter model can explain both the primordial abundance
of dark matter and the anomalous $511~\keV$ gamma-ray signal from
the galactic center.  This model posits a light neutral scalar, $\chi$,
with a mass in the range $1~\MeV < \Mchi < 10~\MeV$, as well 
as a light neutral spin-$1$ boson, $U$, which mediates the annihilation 
channel $\chi\chi^*\rightarrow\epem$.  Since the dark matter particle is
light, its number density is relatively large if it accounts for a
local dark matter density of $\dmdensity$.  We consider an experiment
in which a low-energy, high-current electron beam is passed through
a long evacuated tube, and elastic scattering of electrons off dark
matter particles is observed.  The kinematics of this process allow
a clean separation of the signal process from scattering off residual
gas in the tube, and also a direct measurement of $\Mchi$.
\end{abstract}


\maketitle

\section{\label{S:intro}Introduction}
\par
The nature of dark matter is one of the most interesting questions 
of modern science.  The canonical explanation from particle
physics posits a massive, weakly-interacting particle.  However,
a mass on the order of $100~\GeV$ is not strictly necessary,
and a model based on the idea of {\sl light dark matter particles}
has been proposed by Boehm, Fayet and others~\cite{LDM,Fayet,Ascasibar}.
This model can also explain the anomalous $511~\keV$ gamma-ray signal
from the galactic center observed by INTEGRAL~\cite{INTEGRAL,INTEGRAL2}
and earlier balloon-borne devices~\cite{BALLOON1,BALLOON2}.
\par
The light dark matter model (LDM) posits a light neutral scalar particle,~$\chi$,
with mass in the range $1~\MeV < \Mchi < 10~\MeV$\cite{Beacom}, as well as a light
neutral spin-$1$ boson, $U$, which mediates the annihilation channel
$\chi\chi^*\rightarrow\epem$.   Some versions of the model also include
a heavy charged fermion, $F^\pm$~\cite{Ascasibar}, but this particle
plays no role in the present study.  It is also possible for the light
dark matter particle to be a fermion, but for our present estimates
we will assume that it is a scalar.
\par
Particle physics and astrophysical data constrain the coupling constants
of electrons and of $\chi$ to the $U$-boson as a function of~$\Mchi$ and~$\MU$.
Following Fayet~\cite{Fayet},
\beq
\label{Eq:DMabundance}
  |\Cchi\fe| \approx 10^{-6}
  \frac{M_U^2 - 4M_\chi^2}{\Mchi\,(1.8~\MeV)} \, \sqrt{\Beeann}
\eeq
where $\Cchi$ and $\fe$ are the $U$-$\chi$ and $U$-$e^-$ coupling
constants, respectively, and $\Beeann$ is the fraction of all
$\chi\chi^*$  annihilations which result in an $\epem$ final state.
We take $\Beeann = 1$ in the present study.
\par
If the annihilation channel $\chi\chi^*\rightarrow\epem$ exists, then
the scattering channel $e^-\chi\rightarrow e^-\chi$ also exists with a
cross-section that is {\em directly} related to the annihilation cross-section.
We compute the scattering cross-section on the basis of the~LDM model, and
describe a conceptual experiment to observe this process.  The scattering
rate depends on the local dark matter density, which we take to be $\dmdensity$.
Even with an ultra-high vacuum, a large background from atomic scattering
persists, but this background can be eliminated on the basis of the
kinematics of the scattered electron.  In a previous study~\cite{HKSV}, we
investigated the production of $\chi$-particles in low-energy $e^-\,p$
scattering.  We also computed the contribution of $U$-bosons to
rare pion decay~\cite{KST}.
The present ``evacuated tube'' experiment would confirm
and extend the knowledge gained from the $e^-\,p$-scattering experiment.

\section{\label{S:cs}Scattering Cross Sections}
\par
We consider elastic scattering of relativistic electrons off
quasi-stationary scalar dark matter particles,~$\chi$, via
the $t$-channel exchange of a neutral spin-$1$ boson,~$U$.
We notate the four-vectors as follows:
$p_1$ is the incoming electron, $p_2$ is the $\chi$ particle before collision,
$p_3$ is the outgoing electron, and $p_4$ is the outgoing $\chi$. Let $E$ be the energy
of the incoming electron in the lab frame, and $E'$ be the outgoing energy; as long as
$E \gg m_e$ we can neglect the electron mass and write $p_1 = (E,0,0,E)$ and $p_3 =
(E',E'\sin \theta, 0, E'\cos \theta)$, which defines the electron scattering angle
$\theta$.
\par
We assume that the $U$-boson has a purely vector coupling to the $e^-$, so that the
$U$-$e^-$ vertex factor is $\fe \gamma^{\mu}$. Similarly, the Feynman rule for the
$U$-$\chi$ vertex is $\Cchi(p_4+p_2)^\mu$.
Boehm and Fayet~\cite{LDM} give the amplitude for $\nu$-$\chi$ scattering,
\bdm
|\mathcal{M}|^2 = \frac{\Cchisq \fnusq}{(t-\MUsq)^2} 
  \bigl( (s-u)^2 +t(4 \Mchisq - t) \bigr),
\edm
where $\fnu$ is the $U$-$\nu$ coupling.
This expression coincides with the spin-averaged amplitude for $\chi$-$e^-$
scattering, in the ultra-relativistic limit, with $\fnu$ replaced by $\fe$.
In the lab frame, $t = -4EE'\sin^2(\theta/2)$ and $s-u = 4\Mchi E$, so with our notation,
\begin{eqnarray*}
\langle |\mathcal{M}|^2 \rangle  = \frac{16\Cchisq\fesq}{(4EE' \sin^2(\theta/2) + \MUsq)^2}
\times  \nonumber \\
\left ( \Mchisq E^2 + E^2 E'^2 \sin^4(\theta/2) - EE'\Mchisq \sin^2(\theta/2) \right ).
\end{eqnarray*}
For a $2 \rightarrow 2$ scattering process with a relativistic incident particle, the
differential cross-section is given by
\beq
\frac{d\sigma}{d\Omega} = |\mathcal{M}^2| \left ( \frac{E'}{8\pi \Mchi E} \right )^2.
\label{cs}
\eeq
For $\MU = 10~\MeV$ and $\Mchi = 2~\MeV$, Eq.~(\ref{Eq:DMabundance}) gives
$|\Cchi\fe| \approx 2.3 \times 10^{-5}$. Taking $\Ebeam = 100~\MeV$ and integrating
Eq.~(\ref{cs}) numerically, we obtain a total cross-section of~$138~\pb$.

\section{\label{S:exp}Conceptual Design}
\par
\begin{figure}[htbp]
\begin{center}
\includegraphics[scale=0.3]{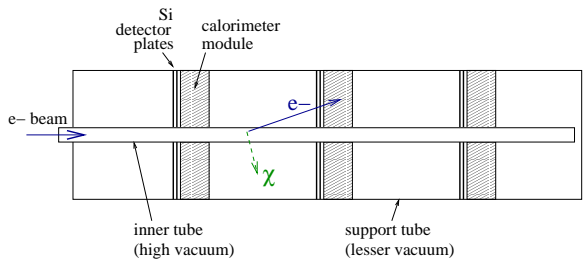}
\end{center}
\caption{\label{f:apparatus} 
drawing of a repeated structure consisting of two Si detector
planes and a calorimeter module, through which the thin tube
at ultra-high vacuum is threaded.  The beam enters at the left.
The outer tube is held at a more ordinary vacuum and supports
the detector modules.  This drawing shows approximately 
one-fiftieth of the entire experiment.  Example trajectories for
a scattered electron and dark matter particle are shown.}
\end{figure}

Our concept for a practical experiment is as follows.  A low-energy,
high-current electron beam is passed through a long evacuated tube.
The energy of the beam is $E = \EbeamVAL$ and the average current is
$\IbeamVAL$. There is no need for the beam to be well-focused,
nor is the time structure important.   The length of the
tube is taken to be $\LtubeVAL$, and its radius, $\RtubeVAL$.  
An ultra-high vacuum is established in the tube, at the level 
of~$\uhvVAL$, which has been achieved already in the laboratory~\cite{UHV}. 
There would be about $3\times 10^{4}$ molecules per cm$^3$ to be compared
to ${\cal{O}}(100)$ dark matter particles.  The ultra-high vacuum tube
is enclosed within a tube of radius $\RouterVAL$, with a lesser
vacuum of $\lessvacVAL$, which contains and supports all the instrumentation.
See Fig.~\ref{f:apparatus} for a drawing.

\par
The kinematics of the outgoing electron would be measured with the
following apparatus.
Two silicon pixel detector plates, spaced $5~\mathrm{mm}$ apart,
would be placed perpendicular to the beam line, followed by a circular array
of calorimeter elements. This arrangement would be repeated every $50~\mathrm{cm}$.
If an electron passes through the pair of detector plates and the calorimeter
array, the two position measurements at the silicon plates would provide a
measurement of~$\theta$, while the calorimeter would measure $E'$. This arrangement 
minimizes the effects of multiple scattering, since the path of the outgoing electron 
is nearly normal to the plane of the plates.
\par
The acceptance of this detector apparatus is limited by the thickness of the
calorimeters, since electrons exiting the inner tube directly under a calorimeter 
array would miss both silicon detectors. To reduce this ``dead area'' it is
advantageous to make the calorimeters as thin as possible while still
providing an energy measurement up to $\EbeamVAL$. Keeping only
$\theta > \ThetaminVAL$ and assuming a calorimeter thickness of $\calthick$, 
we find a geometric acceptance of about $37\%$. 
\par
The signal rate will be far smaller than the background rate, even
for a vacuum of~$\uhvVAL$, because the ``target'' of dark matter
particles is at least a factor hundred thinner than the background,
and the scattering cross section is orders of magnitude smaller.
Hence, the ability to measure $\theta$ and $E'$ accurately is crucial.
\par
Measurement errors on $\theta$ come from multiple scattering and 
the position resolution of the Si plates.  Assuming ultra-thin wafers
with a thickness of $\Sithick$~\cite{Velthuis}, an effective thickness for 
multiple scattering of $0.002 X_0$, and a coordinate measurement error of
$\ErrsiVAL$ on both the $x$- and $y$-coordinates, 
the RMS error on $\theta$ would be in the $1$--$3^\circ$ range.
\par
For the measurement of $E'$, we considered the high-resolution electromagnetic
calorimeter deployed in the BaBar experiment~\cite{BABARCAL}.  This 
state-of-the-art device employs CsI crystals at least 16$X_0$ deep,
and the light is collected by a pair of photo-diodes.  The energy 
resolution is better than 5.1\% for $E' > 50$~MeV.

\section{\label{S:results}Results and Discussion}
\par
\begin{figure}[htbp]
\begin{center}
\includegraphics[scale=0.28]{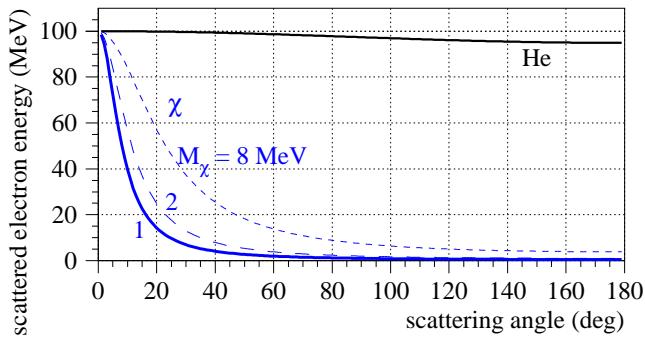}
\end{center}
\caption{\label{F:epvstheta} scattered electron energy~($E'$) vs.
the scattering angle~($\theta$), contrasting the high values for~$E'$
from atomic scattering and the low values from $e^-$-$\chi$ scattering,
for three values of~$\Mchi$}
\end{figure}

We can achieve a clear discrimination of signal and background by
exploiting the relation between scattering angle and outgoing electron energy,
\beq
\label{Eq:Eprimevstheta}
E' = \frac{E}{1+(E/M)(1- \cos \theta)},
\eeq
where $M$ is the mass of the particle off which the electron scatters.
For scattering off a nucleus, $M \gg E$, so $E' \approx E$ for \emph{any}
scattering angle. However, since $\Mchi < E$, $E'$ has a significant
variation with $\theta$, as shown in Fig.~\ref{F:epvstheta}.
Except at zero scattering angle, which is inaccessible in the experiment,
there is a dramatic difference in $E'$ for backgrounds (such as He) and
the signal, for a given~$\theta$.

\par
We sketch a simple analysis as follows.  The apparent scattering
angle is given by the position measurements from the Si plates.  
If we take $\theta > \ThetaminVAL$, then we would expect 
$E' \approx 100$~MeV for background and $E' < 65$~MeV for signal,
depending on $\Mchi$.
Both the signal and background cross sections fall rapidly
with $\theta$, so the event distributions will be concentrated
near $\ThetaminVAL$.  Thus, naively, the $E'$ distribution would
consist of a large narrow peak at $E' \approx E$ for the background,
and a much smaller, broader distribution in the $10 - 70$~MeV range
for signal.  The exact shapes of these distributions will depend 
on the resolution in $\theta$ and $E'$.
A simple estimate based on $\Mchi = 2$~MeV 
gives, for $\theta = 10^\circ$, $E' = 57$~MeV,
$\sigma_\theta = 1^\circ$ and $\sigma_{E'} = 2.5$~MeV.
For $\theta = 45^\circ$, $E' = 6$~MeV, $\sigma_\theta = 3^\circ$
and $\sigma_{E'} = 0.5$~MeV (nominal).
Thus, measurement errors do not significantly distort the
kinematic distribution for the signal, and the success of the
search depends mainly on the rejection power against the background.
\par
The main cut between signal and background comes from the
measurement of~$E'$.  Given the excellent performance
of the BaBar calorimeter, one can expect a rejection power
sufficient to select the signal events from above the tail
of the elastic scattering distribution in~$E'$.  For example,
a cut  $E' < 65$~MeV corresponds to nominally $> 7\sigma$ from
the background peak, and should reject essentially all of the 
background events - the fraction of events remaining assuming 
a simple Gaussian resolution function is incalculably small 
(${\cal{O}}(10^{-26})$).  A firm estimate of the realistic 
rejection power would require detailed prototype studies, 
which is beyond the scope of this paper.
\par
Taking $\theta > \ThetaminVAL$, the ratio of integrated
cross section is $\sigma_{\chi}/\sigma_A = 1.5\times 10^{-8}$,
where $\sigma_\chi$ is $\int_{{\ThetaminVAL}}^{90^\circ} d\sigma$
for $\Mchi = 2$~MeV, and $\sigma_A$ is the corresponding
quantity for a He nucleus.   Thus, a calorimeter which can
distinguish $E' < 65$~MeV from $E' > 99$~MeV at the
level of $10^{-9}$ will be sufficient.
\par
Even if the background can be eliminated, one must obtain
a sufficient signal size to establish discovery and to
measure the properties of the dark matter particle.
With the machine and detector parameters listed in
Section~\ref{S:exp}, for $\Mchi = 2$~MeV and
$\MU = 10$~MeV, and running for $10^7$~s (about 120~days),
the signal yield would be about 53~events.  The yield
drops rapidly with $\Mchi$ but is relatively insensitive
to~$M_U$.
\par
Eq.~(\ref{Eq:Eprimevstheta}) shows that $E'$ at a given~$\theta$
is directly correlated with~$\Mchi$, which allows us to measure
$\Mchi$ given $E$ and measurements of $E'$ and $\theta$.  For
the stated measurements above, we find that the resolution on
$\Mchi$ would be in the $15\% - 20\%$ range, for 
$1 < \Mchi < 8$~MeV.  Thus a sample of just twenty events 
could provide a measurement of $\Mchi$ better than~$5\%$.

\section{Summary}
\par
We described an experiment to find light dark matter
particles in a long evacuated tube by observing scattered
electrons and measuring their angles and energies precisely.
Scattering from residual beam gas is eliminated by virtue
of these kinematics.  For a sufficiently long tube and
intense incoming electron beam, some tens of events could
be collected, allowing a good measurement of the dark
matter particle mass, and of the rate.  This experiment
would be ambitious but not impossible; one would perform
this experiment after an initial observation in elastic
electron-proton scattering as detailed in Ref.~\cite{HKSV}.

\begin{acknowledgments}
We acknowledge the support from the DOE under contract
DE-FG02-91ER40684.

\end{acknowledgments}

\bibliography{evac_tube.bib}

\end{document}